\documentclass[aps,showpacs,nofootinbib,preprintnumbers]{revtex4}
\usepackage{graphicx}
\usepackage{color}
\def \be  {\begin{equation}}
\def \ee  {\end{equation}}
\def \bea {\begin{eqnarray}}
\def \eea {\end{eqnarray}}

\begin{document}

\preprint{ECTP-2011-03}



\title{Fine-Grid Calculations for Stellar Electron and Positron Capture Rates on Fe-Isotopes}
\author{Jameel-Un Nabi}\thanks{Corresponding author} 
\email{jameel@giki.edu.pk}
\affiliation{Faculty of Engineering Sciences, GIK Institute of Engineering Sciences and Technology, Topi 23640, Khyber Pakhtunkhwa, Pakistan}
\affiliation{Egyptian Center for Theoretical Physics (ECTP), MTI University, Al-Mokattam, Cairo, Egypt}

\author{Abdel Nasser Tawfik}
\email{a.tawfik@eng.mti.edu.eg} 
\affiliation{Egyptian Center for Theoretical Physics (ECTP), MTI University, Al-Mokattam, Cairo, Egypt}

\date{\today}

\begin{abstract}
The acquisition of precise and reliable nuclear data is a
prerequisite to success for stellar evolution and nucleosynthesis
studies. Core-collapse simulators find it challenging to generate an
explosion from the collapse of the core of massive stars. It is
believed that a better understanding of the microphysics of
core-collapse can lead to successful results. The weak interaction
processes are able to trigger the collapse and control the lepton-to-baryon
ratio ($Y_{e}$) of the core material. It is suggested that the
temporal variation of $Y_{e}$ within the core of a massive star has
a pivotal role to play in the stellar evolution and a fine-tuning of
this parameter at various stages of presupernova evolution is the
key to generate an explosion. During the presupernova evolution of
massive stars, isotopes of iron, mainly $^{54,55,56}$Fe, are
considered to be key players in controlling $Y_{e}$ ratio via
electron capture on these nuclide. Recently an improved microscopic
calculation of weak interaction mediated rates for iron isotopes was
introduced using the proton-neutron quasiparticle random phase
approximation (pn-QRPA) theory. The pn-QRPA theory allows a
microscopic \textit{state-by-state} calculation of stellar capture
rates which greatly increases the reliability of calculated rates.
The results were suggestive of some fine-tuning of the $Y_{e}$ ratio
during various phases of stellar evolution. Here we present for the
first time the fine-grid calculation of the electron and positron
capture rates on $^{54,55,56}$Fe. Core-collapse simulators may find
this calculation suitable for interpolation purposes and for
necessary incorporation in the stellar evolution codes.

\end{abstract}


\maketitle

\section{Introduction}
During the late phases of stellar evolution of massive stars,
electron capture processes on heavy nuclei dominate the time
evolution of the lepton-to-baryon ratio ($Y_{e}$) of the core
material primarily due to the low entropy of the stellar core but
also due to the resulting dominance of heavy nuclei over free
nucleons \cite{Bet79}. Inside the iron core of massive stars, the
electron capture occurs via the Gamow-Teller, GT$_{+}$, transitions
changing protons in the $^{1}f_{7/2}$ level into neutrons in the
$^{1}f_{5/2}$ levels. On the other hand the GT$_{-}$ strength is
responsible for transforming a neutron into a proton (resulting in
positron capture processes). It is conjectured that the electron
captures on proton and positron captures on neutron play a very
crucial role in the supernovae dynamics \cite{Nab99a}. Due to the
weak interaction processes, mainly electron and positron captures
onto nuclei and $\beta^{\pm}$-decays of nuclei, the value of $Y_{e}$
for a massive star changes from $1$ (during hydrogen burning) to
roughly $0.5$ (at the beginning of carbon burning)  and finally to
around $0.42$ just before the collapse leading to a supernova
explosion \cite{Pri00}. The temporal variation of $Y_{e}$ within the
core of a massive star has a pivotal role to play in the stellar
evolution and a fine-tuning of this parameter at various stages of
presupernova evolution is the key to generate an explosion
\cite{Kot06}. Furthermore, the electron capture rates is assumed to
determine the mass of the core and thus the fate of the shock wave
formed by the supernova explosion \cite{Kot06}. The calculation of
electron and positron capture rates is very sensitive to the
distribution of the GT$_{\pm}$ strength functions \cite{Roe93}.
During the early phases of presupernova evolution, the electron
chemical potential is of the same order of magnitude as the nuclear
$Q$ value and the capture rates are sensitive to the details of the
GT$_{\pm}$ strength functions in daughter nuclei \cite{Auf96}. With
proceeding collapse the stellar density increases by orders of
magnitude and the resulting electron chemical potential is
significantly higher than the nuclear $Q$ value and then the capture
rates are largely determined by the centroid and the total strength
of the GT$_{\pm}$ strength functions. A microscopic calculation of
ground and excited state GT$_{\pm}$ strength functions is the key to
reliably estimate the stellar capture rate \cite{Auf96}. The
positron captures are of great importance at high temperature and
low density locations \cite{Ful82}. In such conditions, a rather
high concentration of positron can be reached from $e^{-} +e^{+}
\leftrightarrow \gamma +\gamma $ equilibrium favoring the $e^{-}
e^{+} $ pairs. The electron captures on proton and positron captures
on neutron are considered important ingredients in the modeling of
Type II supernovae \cite{Nab99a}.

The pn-QRPA theory \cite{Hal67} is an efficient way to
microscopically generate GT$_{\pm}$ strength functions which
constitute a primary and nontrivial contribution to the weak
interaction mediated rates among iron-regime nuclide. In this model
a quasiparticle basis is first constructed with a pairing
interaction. The random phase approximation (RPA) equation is then
solved with a schematic GT residual interaction. The pn-QRPA model
has been developed by Halbleib and Sorensen \cite{Hal67} by
generalizing the usual RPA to describe charge-changing transitions
of the type $(Z,N) \rightarrow (Z \pm 1, N \mp 1)$. The model is
then extended to deformed nuclei (using Nilsson wave functions
\cite{Nil55}) by Krumlinde and M\"{o}ller \cite{Kru84}. Extension of
the model to treat odd-odd nuclei and transitions from nuclear
excited states has been done by Muto and collaborators \cite{Mut92}.
The pn-QRPA is a simple and microscopic model. The pn-QRPA model has
two big advantages for performing weak interaction calculations in
stellar matter. Firstly it can handle any arbitrarily heavy system
of nucleons since the calculation is performed in a luxurious model
space of up to $7$ major oscillator shells. The second advantage is
even more important for the calculation of weak interaction rates in
stellar environment. The prevailing temperature of the stellar
matter is of the order of a few hundred kilo-electron volts to a few
million-electron volts and GT$_{\pm}$ transitions occur not only
from nuclear ground state, but also from excited states. As
experimental information about excited state strength functions
seems inaccessible, Aufderheide \cite{Auf91} stressed much earlier
the need to probe these strength functions theoretically. Today the
pn-QRPA theory can be utilized to calculate GT$_{\pm}$ strength
distribution of \textit{all} excited states of parent nucleus in a
microscopic fashion and this feature of the pn-QRPA model greatly
enhances the reliability of the calculated rates in stellar matter
\cite{Nab04}. In other words, the pn-QRPA model seems to allow a
microscopic state-by-state calculation of all stellar weak rates and
the Brink hypothesis is not assumed in this model. Brink hypothesis
\cite{Bri58} states that GT strength distribution on excited states
is \textit{identical} to that from ground state, shifted
\textit{only} by the excitation energy of the state. Recent
calculations have pointed towards the fact that Brink hypothesis is
not a safe approximation to use for calculation of stellar weak
interaction rates \cite{Nab08, Nab09, Nab10, Nab10a, Nab10b}.

Recently, weak interaction rates of isotopes of iron,
$^{54,55,56}$Fe, have been calculated using the pn-QRPA model
\cite{Nab09}. There, the author reported that the calculated
electron capture rates on these iron isotopes were overall in fair
agreement with the large-scale shell model (LSSM) results
\cite{Lan00}. However, it has been found that the calculated beta
decay rates are suppressed by three to five orders of magnitude. In
present paper, we present the fine-grid calculation of stellar
electron and positron capture rates for iron-isotopes
$^{54,55,56}$Fe using pn-QRPA model. Recently, the improved pn-QRPA
calculation has been introduced \cite{Nab09}, where it was reported
that the betterment resulted mainly from the incorporation of
measured deformation values for these nuclei. A detailed analysis of
the calculated ground and excited state GT$_{\pm}$ strength
distributions for $^{54,55,56}$Fe and the capture rates has been
presented in Ref. \cite{Nab10c}. Due to the extreme conditions
prevailing in the cores of massive stars, interpolation of
calculated rates within large intervals of temperature-density
points posed some uncertainty in the values of capture rates for
collapse simulators \cite{Ful82}. As such the calculation is done on
a detailed temperature and density grid pertinent to presupernova
and supernova environment and should prove more suitable for running
on simulation codes. In present work, a detailed calculation of
electron and positron capture rates on $^{54,55,56}$Fe is being
presented for the first time at temperature-density intervals
suitable for simulation purposes. Further, we also compare the
pn-QRPA capture rates with previous key calculations for the
astrophysically important range of stellar temperatures and
densities. It is worthwhile to mention that the presence of
$^{54,55,56}$Fe in the stellar core is a result of the last phase of
silicon burning.

The paper is organized as follows. Section \ref{sec:2} briefly
discusses the formalism of the pn-QRPA calculations. In section
\ref{sec:3}, the calculated results are present. Comparison with
earlier calculations during presupernova evolution of massive stars
is also included in this section. We summarize the main conclusions
in Section \ref{sec:4}. The fine-grid calculation of the stellar
electron and positron capture rates on $^{54,55,56}$Fe is presented
in Table~\ref{tab.1}.

\section{Nuclear model and calculation}
\label{sec:2}
The electron capture (ec) and positron capture (pc) rates of a
transition from the $i$-th state of the parent to the $j$-th state of
the daughter nucleus is given by
\begin{eqnarray} \label{eq:1a}
\lambda ^{^{ec(pc)} } _{ij} &=& \left[\frac{\ln 2}{D}
\right]\left[f_{ij} (T,\rho ,E_{f} )\right]\left[B(F)_{ij}
+\left({\raise0.7ex\hbox{$ g_{A}  $}\!\mathord{\left/ {\vphantom
{g_{A}  g_{V} }} \right.
\kern-\nulldelimiterspace}\!\lower0.7ex\hbox{$ g_{V}  $}}
\right)^{2} B(GT)_{ij} \right].
\end{eqnarray}
The value of D is reported in Ref. \cite{Yos88}. $B_{ij}$ are the
sum of reduced transition probabilities of the Fermi B(F) and GT
transitions B(GT). Whereas for $^{54,56}$Fe phonon transitions
contribute, in the case of $^{55}$Fe (odd-A case) two kinds of
transitions are possible within the framework of the pn-QRPA model.
One are the phonon transitions, where the odd quasiparticle acts as
spectator and the other is the transitions of the odd quasiparticle
itself. In the later case phonon correlations have been introduced
to one-quasiparticle states in first-order perturbation
\cite{Mut89}. The functions $f_{ij}$ give the phase space integrals.
Details of the calculations of phase space integrals and reduced
transition probabilities in the pn-QRPA model can be found in Refs.
\cite{Nab10b, Nab10c}.

Therefore, the total electron (positron) capture rate per unit time
per nucleus reads
\begin{eqnarray} \label{eq:2a}
\lambda^{ec(pc)} &=& \sum _{ij}P_{i} \lambda _{ij}^{ec(pc)}.
\end{eqnarray}
The summation over all initial and final states is to be carried out
until satisfactory convergence in the rate calculations is achieved.
Here $P_{i}$ is the probability of occupation of parent excited
states and follows the normal Boltzmann distribution. The pn-QRPA
theory allows a microscopic state-by-state calculation of both sums
present in Eq. (\ref{eq:2a}). As discussed earlier in previous
section, this feature of the pn-QRPA model greatly increases the
reliability of the calculated rates in stellar matter where there
exists a finite probability of occupation of excited states.

In order to further increase the reliability of the calculated
capture rates experimental data were incorporated in the calculation
wherever possible. In addition to the incorporation of the
experimentally adopted value of the deformation parameter, the
calculated excitation energies (along with their log $ft$ values)
were replaced with an experimental one when they were within 0.5 MeV
of each other. The calculation presented in this work takes into
consideration certain aspects. The missing measured states are
inserted and inverse and mirror transitions are also taken into
account. No theoretical levels have been replaced with the
experimental ones beyond the excitation energy for which
experimental compilations had no definite spin and/or parity. A
state-by-state calculation of GT$_{\pm}$ strength has been performed
for a total of $246$ parent excited states in $^{54}$Fe, $297$
states for $^{55}$Fe and $266$ states for $^{56}$Fe. For each parent
excited state, transitions are calculated to $150$ daughter excited
states. The band widths of energy states are chosen according to the
density of states to cover an excitation energy of ($15-20$) MeV in
parent and daughter nuclei. The summation in Eq. (\ref{eq:2a}) has
been done to ensure satisfactory convergence. The use of a separable
interaction assisted in the incorporation of a luxurious model space
of up to $7$ major oscillator shells which in turn made possible to
consider these many excited states both in parent and daughter
nucleus.

Recently, the deformation parameter is being argued as an important
parameter for QRPA calculations {\it at par} with the pairing parameter by
Stetcu and Johnson \cite{Ste04}. As such rather than using
deformations from some theoretical mass model (as used in earlier
calculations of pn-QRPA rates e.g. \cite{Nab99, Nab04}), the
experimentally adopted value of the deformation parameters for
$^{54,56}$Fe, extracted by relating the measured energy of the first
$2^{+}$ excited state with the quadrupole deformation, has been taken
from Raman {\it et al.} \cite{Ram87}. For the case of $^{55}$Fe, where
such measurement apparently lacks, the deformation of the nucleus has
 been calculated as
\begin{eqnarray} \label{eq:3a}
\delta &=& \frac{125(Q_{2})}{1.44 (Z) (A)^{2/3}},
\end{eqnarray}
where $Z$ and $A$ are the atomic and mass numbers, respectively.
$Q_{2}$ is the electric quadrupole moment taken from Ref.
\cite{Moe81}. Q-values are taken from the recent mass compilation
of Audi {\it et al.} \cite{Aud03}.

The incorporation of measured deformations for $^{54,56}$Fe and a
smart choice of strength parameters led to an improvement of the
calculated GT$_{\pm}$ distributions compared to the measured ones
\cite{Nab09}. In Table 1 of Ref. \cite{Nab09}, it has been shown
that the present pn-QRPA calculated GT$_{\pm}$ centroids and total
$S_{\beta^{\pm}}$ strengths were in good agreement with available
data for the even-even isotopes of iron. The table also shows marked
improvement in the reported pn-QRPA calculation over the previous
one \cite{Nab04}.

The fine-grid calculation of stellar electron and positron capture
rates on $^{54,55,56}$Fe as a function of stellar temperature and
density is given in Table~\ref{tab.1}. The calculated rates are
tabulated in logarithmic (to base $10$) scale. It is simply a matter
of convention to give $-100$ for instance to express that the rate
is smaller than $10^{-100}$. The first column gives log($\rho
Y_{e}$) in units of g$\,$ cm$^{-3}$, where $\rho$ is the baryon
density and $Y_{e}$ is the ratio of the electron number to the
baryon number. Stellar temperatures ($T_{9}$) are given in $10^{9}$
K. Stated also are the values of the Fermi energy of electrons in
units of MeV. $\lambda^{ec}$($\lambda^{pc}$) are the electron
(positron) capture rates in units of  $sec^{-1}$, Eq. (\ref{eq:2a}).
Core-collapse simulators may find it useful to employ the reported
electron and positron capture rates on iron isotopes in their codes.
The ASCII file of Table~\ref{tab.1} is also available and can be
received from the corresponding author upon request.

A detailed comparison of reported capture rates with previous
calculations over a wide range of temperature and density has
 been presented earlier \cite{Nab10c}. Here, we would like to present how
the reported rates compare with previous calculations during
temperature and density domains of astrophysical interest. For the
sake of comparison, we took into consideration the pioneer
calculations of FFN \cite{Ful82}, those performed using the
large-scale shell model (LSSM) \cite{Lan00} and the recently
reported thermal QRPA (TQRPA) approach \cite{Dzh10}.

An analysis of the contribution of excited states to the total
capture rates for these iron isotopes was performed earlier by Nabi
\cite{Nab10c}. There, it was shown that the electron capture rates
have a significant contribution from parent excited states for the
case of $^{54}$Fe.  The total capture rate is dominated by ground
state contribution alone for the case of $^{55}$Fe and the case of
$^{56}$Fe had a $50$-$50$ contribution. Whenever ground state and
low-lying states ($\sim$ 1 MeV) dominate the total capture rates the
microscopic calculations of pn-QRPA and LSSM  are found to be in
very good agreement.

\section{Results and comparisons}
\label{sec:3} 
\begin{figure}[htbp]
\includegraphics[width=10cm]{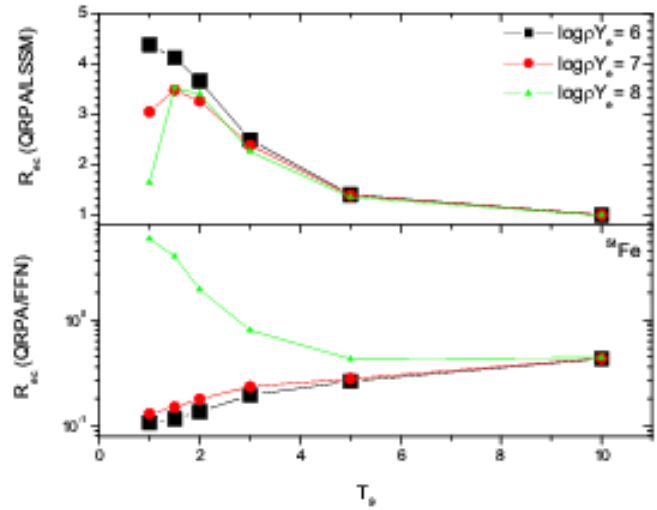}
\caption{(Color online) Ratios of reported electron capture rates to
those calculated using LSSM \cite{Lan00} (upper panel) and by FFN
\cite{Ful82} (lower panel) on $^{54}$Fe as function of stellar
temperatures and densities. T$_{9}$ gives the stellar temperature in
units of $10^{9}$ K. In the legend, $log \rho Y_{e}$ gives the log
to base 10 of stellar density in units of $gcm^{-3}$.}
\label{figure1}
\end{figure}

Figure~\ref{figure1} depicts the comparison of
electron capture rates on $^{54}$Fe with LSSM and FFN calculations.
The graph is drawn for temperature and density domain of
astrophysical importance (oxygen shell burning and silicon core
burning phases) for the case of $^{54}$Fe. The upper panel displays
the ratio of calculated rates to the LSSM rates,
$R_{ec}$(QRPA/LSSM), while the lower panel shows a similar
comparison with the FFN calculation, $R_{ec}$(QRPA/FFN). At lower
stellar temperatures and densities the calculated electron capture
rates are bigger than the corresponding LSSM rates by as much as a
factor of four. The results agree at high temperatures and
densities. During the early stages of the collapse the capture rates
are very sensitive to the  excited GT$_{+}$ strength distributions.
It is reminded that LSSM approach allows for detailed nuclear
spectroscopy but partially employs the Brink hypothesis. The pn-QRPA
model, on the other hand, performs a microscopic calculation of all
low-lying GT$_{+}$ strength distributions. At high densities the
capture rates depend more on the centroid and total strength of the
GT$_{+}$ functions. It was shown in Table 1 of Ref. \cite{Nab09}
that the pn-QRPA calculated the centroid at a slightly higher energy
in daughter $^{54}$Mn nucleus as against LSSM centroid. But the
effect was compensated by the calculation of a much higher total
strength value as compared to LSSM strength. Accordingly the two
results match well at high densities. The FFN electron capture rates
are on the average bigger by an order of magnitude (lower panel of
Figure~\ref{figure1}) as compared to pn-QRPA rates. There are two
main reasons for this enhancement in FFN rates. Firstly FFN did not
take into consideration the quenching of the GT strength. FFN also
did not take into effect the process of particle emission from
excited states and accordingly their parent excitation energies
extended well beyond the particle decay channel. For these reasons
FFN rates were also bigger than LSSM rates by around an order of
magnitude.

\begin{figure}[htbp]
\includegraphics[width=10cm]{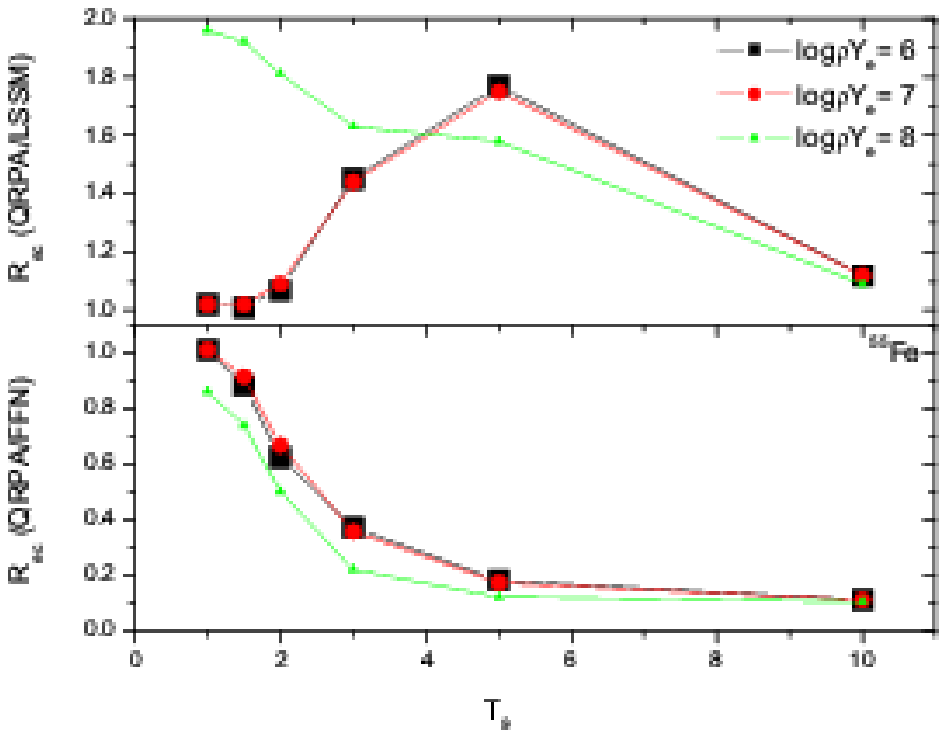}
\caption{(Color online) Same as figure~\ref{figure1} but for
electron capture rates on $^{55}$Fe.}\label{figure2}
\end{figure}

The electron capture rates on $^{55}$Fe are argued to be most
effective during the oxygen shell burning till around the ignition
of the first stage of convective silicon shell burning of massive
stars. Figure~\ref{figure2} depicts the comparison of the calculated
electron capture rates on $^{55}$Fe with LSSM (upper panel) and FFN
(lower panel) calculations for the corresponding physical
conditions. The upper panel shows a very good comparison of pn-QRPA
and LSSM rates (within a factor of $2$). Primarily the ground state
but also a few low-lying excited GT$_{+}$ strength distributions
play a key role in the calculation of total capture rates for this
odd-A nucleus. LSSM performs a microscopic calculation of these
low-lying strength functions (up to around $1$ MeV). As mentioned
earlier the ground state GT$_{+}$ strength function dominates the
total capture rates on $^{55}$Fe and the two calculations agree very
well. The FFN rates are again bigger by an order of magnitude at
high temperatures. The probability of occupation of high-lying
excited states increases with stellar temperature and FFN rates are
much bigger for reasons given above.

\begin{figure}[htbp]
\includegraphics[width=10cm]{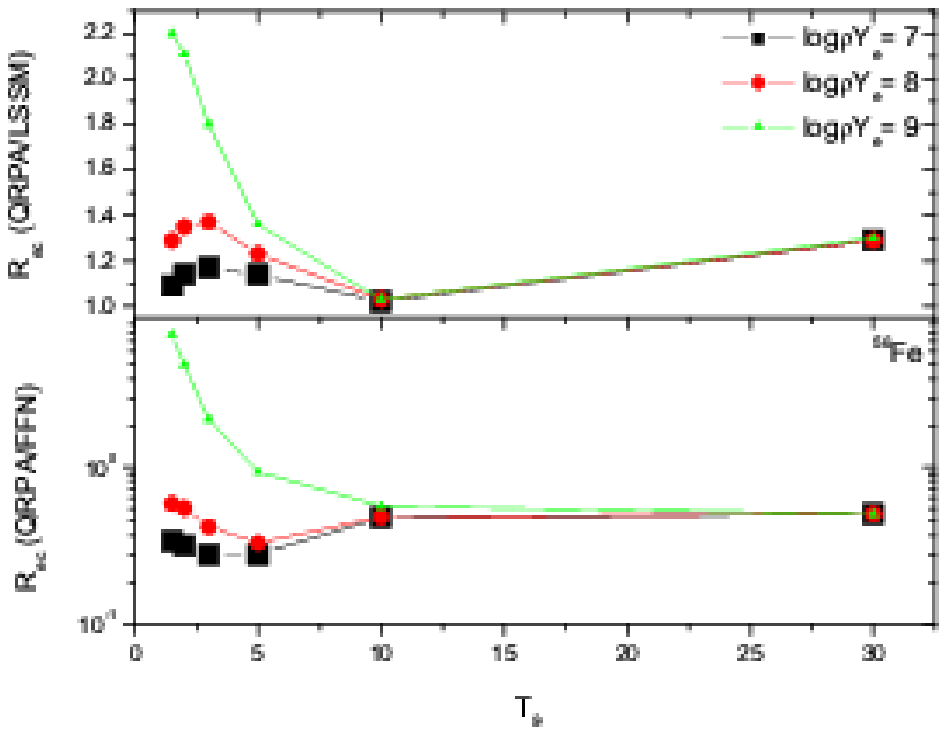}
\caption{(Color online) Same as figure~\ref{figure1} but for
electron capture rates on $^{56}$Fe.}\label{figure3}
\end{figure}

The comparison of electron capture rates on $^{56}$Fe with previous
calculations is shown in Figure~\ref{figure3}. The capture rates on
this isotope are very important for the pre-supernova phase of
massive stars. As such Figure~\ref{figure3} shows the comparison up
to $T_{9} [K] = 30$ and $\rho\, Y_{e} [g\,cm^{-3}] =10^{9}$. Once
again the comparison between pn-QRPA and LSSM is very good (for
reasons mentioned earlier). FFN rates are again bigger for obvious
reasons.

\begin{figure}[htbp]
\includegraphics[width=10cm]{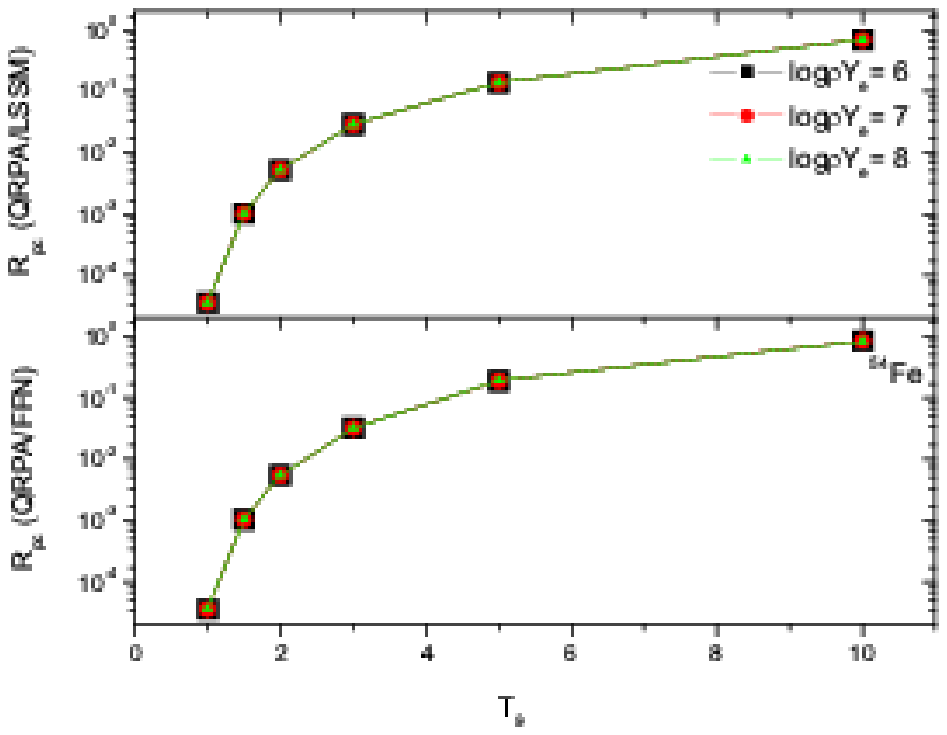}
\caption{(Color online) Same as figure~\ref{figure1} but for
positron capture rates on $^{54}$Fe.}\label{figure4}
\end{figure}

There are two key facts to keep in mind before  we present a
comparison of positron capture rates for $^{54,55,56}$Fe. First, of
all it is to be noted that positron capture rates are tens of orders
of magnitude smaller than the corresponding electron capture rates.
Secondly, these small numbers are fragile functions of the available
phase space and can change appreciably by a mere change of $0.5$ MeV
in phase space calculations. The positron capture rates are more
indicative of the uncertainties in calculation of the energy
eigenvalues (for both parent and daughter states). Further it was
shown in Ref. \cite{Nab10c} that the ground state rate contributed
at the maximum by only about $33\%$ to the total positron capture
rates. Figure~\ref{figure4} shows the comparison of the calculated
positron capture rates on $^{54}$Fe with the LSSM and FFN
calculation for similar physical conditions as depicted in
Figure~\ref{figure1}. Here one sees that the pn-QRPA rates are
suppressed by up to five orders of magnitude at low temperatures.
The comparison improves as temperature increases. The comparison is
similar in both upper and lower panels and does not change by
increasing density by two orders of magnitude (from $\rho Y_{e}
[gcm^{-3}] =10^{6}$ to $\rho\, Y_{e} [g\,cm^{-3}] =10^{8}$).

\begin{figure}[htbp]
\includegraphics[width=10cm]{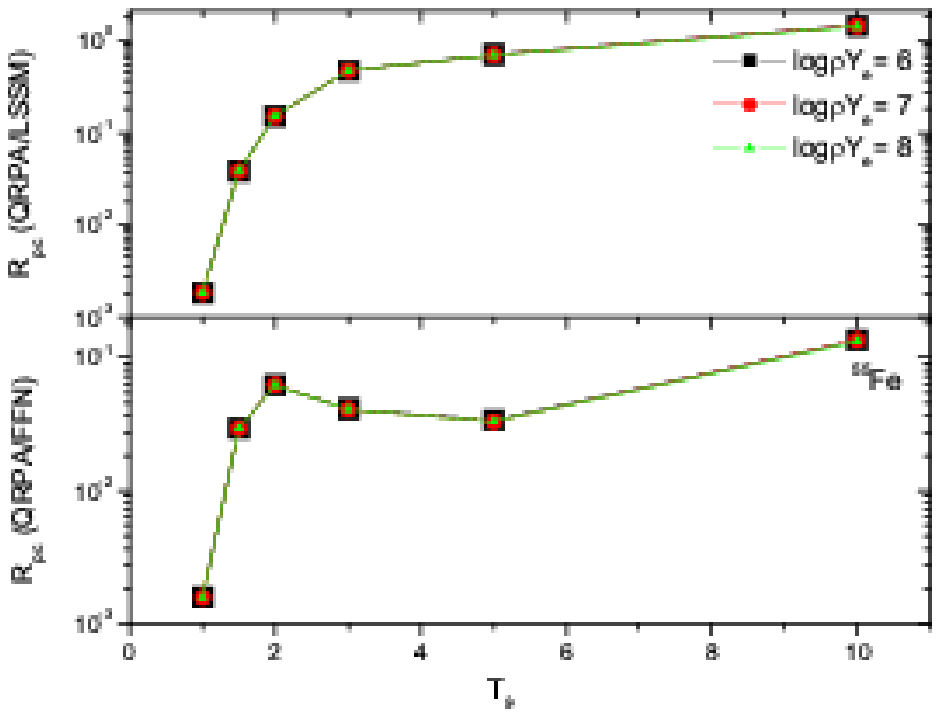}
\caption{(Color online) Same as figure~\ref{figure1} but for
positron capture rates on $^{55}$Fe.}\label{figure5}
\end{figure}

\begin{figure}[htbp]
\includegraphics[width=10cm]{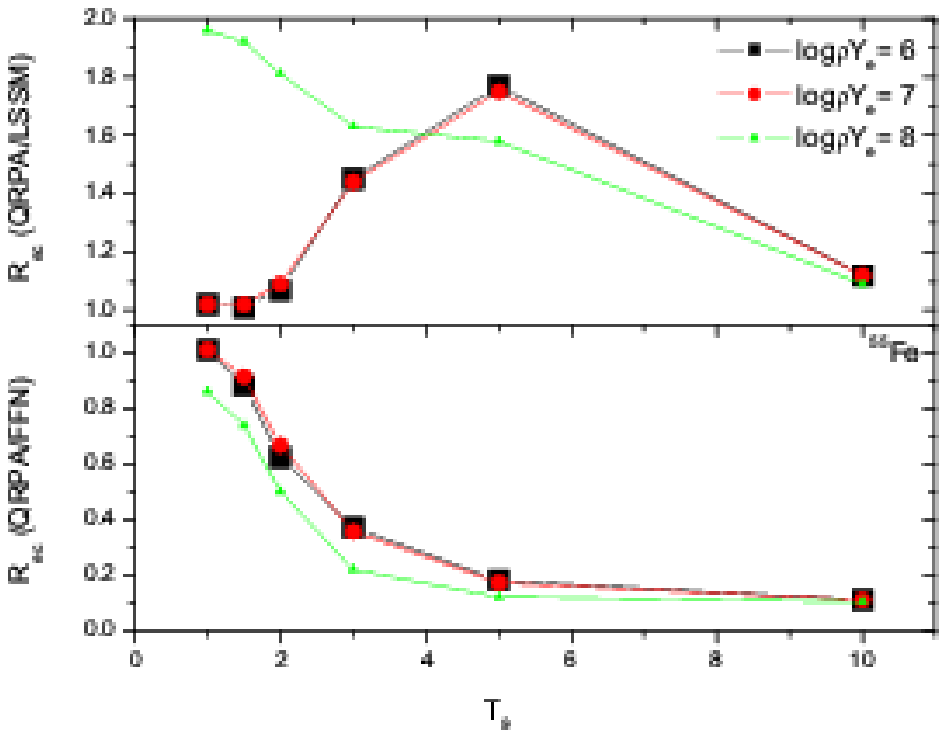}
\caption{(Color online) Same as figure~\ref{figure1} but for
positron capture rates on $^{56}$Fe.}\label{figure6}
\end{figure}

The comparison between pn-QRPA and LSSM improves for the case of
$^{55}$Fe (Figure~\ref{figure5}) where the rates differ at the
maximum by three orders of magnitude. The positron capture rates are
in very good agreement with LSSM rates at high temperatures. Still
better is the comparison case for  $^{56}$Fe (Figure~\ref{figure6}).
The FFN rates are bigger by up to four orders of magnitude.

Recently Dzhioev and collaborators \cite{Dzh10} applied the pn-QRPA
model extended to finite temperature by the thermofield dynamics
formalism (referred to as TQRPA) and calculated electron capture
rates on $^{54,56}$Fe. It was reported by the authors that TQRPA and
LSSM results matched well at low temperatures and high densities. At
high temperatures the TQRPA rates were bigger than LSSM electron
capture rates. Table~\ref{tab.2} presents a mutual comparison
between pn-QRPA, TQRPA and LSSM electron capture rates on these
even-even isotopes of iron. It can be seen from Table~\ref{tab.2}
that TQRPA electron capture rates on $^{54}$Fe are smaller by two
(six) orders of magnitude at $T_{9} [K] = 1$ and $\rho\, Y_{e}
[g\,cm^{-3}] =10^{7} (10^{8})$ as compared to pn-QRPA and LSSM
rates. The comparison improves as temperature and density increase
and TQRPA rates are doubled at $T_{9} [K] = 10$. As far as
comparison of electron capture rates on $^{56}$Fe is concerned, one
again notes that TQRPA rates are smaller by up to seven orders of
magnitude at low temperatures and densities. However by merely
changing $T_{9} [K] = 1$ to $T_{9} [K] = 1.5$ the TQRPA rates gets
up to $4$ orders of magnitude bigger than the corresponding pn-QRPA
and LSSM rates. It looks that TQRPA rates are very sensitive to
temperature changes for the case of $^{56}$Fe at low densities. The
comparison improves with increasing stellar temperatures and
densities. Once again at $T_{9} [K] = 10$ the TQRPA rates are
roughly double the pn-QRPA and LSSM rates. Table~\ref{tab.2} shows
that the comparison between pn-QRPA and LSSM is comparatively far
better and that TQRPA model requires further refinement.

\section{Conclusions}
\label{sec:4} Precise and reliable nuclear data can lead to
deciphering the elemental and isotopic abundances in Nature with
more confidence and thereby derive meaningful constraints on the
associated astrophysical models. The pn-QRPA model has a good track
record of calculation of weak interaction rates both in terrestrial
and stellar domains. The model has access to a huge model space
making it possible to calculate weak rates for arbitrarily heavy
system of nucleons. Further the model gets rid of the Brink
hypothesis and calculates a \textit{state-by-state} calculation of
stellar capture rates which greatly increases the reliability of
calculated rates. Incorporation of experimental deformation lead to
a much improved version of this calculation. The model was used
recently to calculate weak interaction mediated rates on iron
isotopes, $^{54,55,56}$Fe \cite{Nab09}. Electron capture on these
isotopes of iron are mainly responsible for decreasing the
electron-to-baryon ratio during the oxygen and silicon burning
phases of massive stars. The corresponding positron capture rates
are orders of magnitude smaller and accordingly less important for
late stellar evolution of massive stars.

The calculated electron and positron capture rates are also compared
against previous key calculations of weak rates. During the oxygen
and silicon core burning phase of massive stars, the pn-QRPA
electron capture rates on $^{54}$Fe are around three times bigger
than those calculated by LSSM. The comparison is very good for the
case of $^{55,56}$Fe. FFN electron capture rates are bigger by an
order of magnitude. Orders of magnitude differences in electron
capture rates exist between TQRPA and LSSM/pn-QRPA model.

The former calculations of positron capture rates on $^{54,55,56}$Fe
are $3$-$5$ orders of magnitude bigger than the pn-QRPA rates. The
LSSM positron capture rates on $^{56}$Fe is in good agreement with
reported rates.

The main idea of reporting this work is to present a fine-grid
calculation of electron and positron capture rates on
$^{54,55,56}$Fe in stellar matter. Such tables (e.g. Table II of
Ref. \cite{Nab08a}) are of great utility for core-collapse
simulators and also suitable for interpolation purposes. The
fine-grid calculation is presented in Table~\ref{tab.1}. The ASCII
file of this table can also be requested from the author.
Core-collapse simulators are encouraged to employ these rates in
simulation codes to check for possible interesting outcomes.

\section*{Acknowledgment}
J.-U. Nabi would like to acknowledge the support provided by
TWAS and the local hospitality provided by the Egyptian Center for
Theoretical Physics (ECTP), Cairo-Egypt, where this project was
partially completed. J.-U. Nabi also wishes to acknowledge the
support of research grant provided by the Higher Education
Commission, Pakistan, through HEC Project No. 20-1283.

\clearpage

\begin{table}
\caption {Calculated electron and positron capture rates on
$^{54,55,56}$Fe on a fine-grid temperature-density scale.  The
calculated rates are tabulated in logarithmic (to base 10) scale. In
the table, -100 means that the rate is smaller than 10$^{-100}$. For
units see text.\label{tab.1}}
\begin{center}
\footnotesize
\begin{tabular}{|ccc|cccccc|} \hline\hline
$log\rho Y_{e}$ & $T_{9}$ & $E_{f}$ & \multicolumn{2}{|c|}{$^{54}$Fe}& \multicolumn{2}{|c|}{$^{55}$Fe}
& \multicolumn{2}{|c|}{$^{56}$Fe} \\
\cline{4-9}
& & & \multicolumn{1}{c}{$\lambda_{ec}$} &             \multicolumn{1}{|c|}{$\lambda_{pc}$} &
\multicolumn{1}{c}{$\lambda_{ec}$} &                   \multicolumn{1}{|c|}{$\lambda_{pc}$} &
\multicolumn{1}{c}{$\lambda_{ec}$} &                   \multicolumn{1}{|c|}{$\lambda_{pc}$} \\ \hline
   0.5 &   0.50 &   0.065 &    -24.273 &    -92.182 &    -11.945 &    -44.428 &    -47.582 &    -55.832   \\
   0.5 &   1.00 &   0.000 &    -14.748 &    -46.930 &     -9.415 &    -24.160 &    -25.753 &    -29.032   \\
   0.5 &   1.50 &   0.000 &    -11.123 &    -31.923 &     -8.077 &    -17.296 &    -17.897 &    -20.127   \\
   0.5 &   2.00 &   0.000 &     -9.119 &    -24.319 &     -7.159 &    -13.784 &    -13.821 &    -15.584   \\
   0.5 &   2.50 &   0.000 &     -7.758 &    -19.668 &     -6.382 &    -11.623 &    -11.293 &    -12.799   \\
   0.5 &   3.00 &   0.000 &     -6.738 &    -16.496 &     -5.716 &    -10.136 &     -9.554 &    -10.895   \\
\end{tabular}
\end{center}
\end{table}
\begin{table}
\begin{center}
\footnotesize\begin{tabular}{|ccc|cccccc|} \hline\hline
$log\rho Y_{e}$ & $T_{9}$ & $E_{f}$ & \multicolumn{2}{|c|}{$^{54}$Fe}& \multicolumn{2}{|c|}{$^{55}$Fe}
& \multicolumn{2}{|c|}{$^{56}$Fe} \\
\cline{4-9}
& & & \multicolumn{1}{c}{$\lambda_{ec}$} &             \multicolumn{1}{|c|}{$\lambda_{pc}$} &
\multicolumn{1}{c}{$\lambda_{ec}$} &                   \multicolumn{1}{|c|}{$\lambda_{pc}$} &
\multicolumn{1}{c}{$\lambda_{ec}$} &                   \multicolumn{1}{|c|}{$\lambda_{pc}$} \\ \hline
   1.5 &   0.50 &   0.162 &    -23.295 &    -93.160 &    -10.966 &    -45.406 &    -46.604 &    -56.810   \\
   1.5 &   1.00 &   0.002 &    -14.741 &    -46.936 &     -9.408 &    -24.166 &    -25.746 &    -29.038   \\
   1.5 &   1.50 &   0.000 &    -11.122 &    -31.923 &     -8.077 &    -17.297 &    -17.896 &    -20.128   \\
   1.5 &   2.00 &   0.000 &     -9.119 &    -24.319 &     -7.159 &    -13.784 &    -13.820 &    -15.584   \\
   1.5 &   2.50 &   0.000 &     -7.758 &    -19.668 &     -6.381 &    -11.623 &    -11.292 &    -12.799   \\
   1.5 &   3.00 &   0.000 &     -6.738 &    -16.496 &     -5.716 &    -10.136 &     -9.553 &    -10.895   \\
\end{tabular}
\end{center}
\end{table}
\begin{table}
\begin{center}
\footnotesize\begin{tabular}{|ccc|cccccc|} \hline\hline
$log\rho Y_{e}$ & $T_{9}$ & $E_{f}$ & \multicolumn{2}{|c|}{$^{54}$Fe}& \multicolumn{2}{|c|}{$^{55}$Fe}
& \multicolumn{2}{|c|}{$^{56}$Fe} \\
\cline{4-9}
& & & \multicolumn{1}{c}{$\lambda_{ec}$} &             \multicolumn{1}{|c|}{$\lambda_{pc}$} &
\multicolumn{1}{c}{$\lambda_{ec}$} &                   \multicolumn{1}{|c|}{$\lambda_{pc}$} &
\multicolumn{1}{c}{$\lambda_{ec}$} &                   \multicolumn{1}{|c|}{$\lambda_{pc}$} \\ \hline
   2.5 &   0.50 &   0.261 &    -22.295 &    -94.160 &     -9.966 &    -46.407 &    -45.604 &    -57.810   \\
   2.5 &   1.00 &   0.015 &    -14.673 &    -47.005 &     -9.340 &    -24.235 &    -25.678 &    -29.106   \\
   2.5 &   1.50 &   0.002 &    -11.117 &    -31.928 &     -8.072 &    -17.301 &    -17.892 &    -20.132   \\
   2.5 &   2.00 &   0.000 &     -9.118 &    -24.320 &     -7.158 &    -13.785 &    -13.819 &    -15.585   \\
   2.5 &   2.50 &   0.000 &     -7.758 &    -19.668 &     -6.381 &    -11.623 &    -11.292 &    -12.799   \\
   2.5 &   3.00 &   0.000 &     -6.738 &    -16.496 &     -5.716 &    -10.136 &     -9.553 &    -10.895   \\
\end{tabular}
\end{center}
\end{table}
\begin{table}
\begin{center}
\footnotesize\begin{tabular}{|ccc|cccccc|} \hline\hline
$log\rho Y_{e}$ & $T_{9}$ & $E_{f}$ & \multicolumn{2}{|c|}{$^{54}$Fe}& \multicolumn{2}{|c|}{$^{55}$Fe}
& \multicolumn{2}{|c|}{$^{56}$Fe} \\
\cline{4-9}
& & & \multicolumn{1}{c}{$\lambda_{ec}$} &             \multicolumn{1}{|c|}{$\lambda_{pc}$} &
\multicolumn{1}{c}{$\lambda_{ec}$} &                   \multicolumn{1}{|c|}{$\lambda_{pc}$} &
\multicolumn{1}{c}{$\lambda_{ec}$} &                   \multicolumn{1}{|c|}{$\lambda_{pc}$} \\ \hline
   3.5 &   0.50 &   0.361 &    -21.296 &    -95.164 &     -8.967 &    -47.411 &    -44.600 &    -58.814   \\
   3.5 &   1.00 &   0.115 &    -14.172 &    -47.507 &     -8.839 &    -24.737 &    -25.176 &    -29.608   \\
   3.5 &   1.50 &   0.015 &    -11.072 &    -31.974 &     -8.026 &    -17.347 &    -17.846 &    -20.178   \\
   3.5 &   2.00 &   0.004 &     -9.108 &    -24.330 &     -7.148 &    -13.795 &    -13.809 &    -15.595   \\
   3.5 &   2.50 &   0.002 &     -7.754 &    -19.672 &     -6.377 &    -11.627 &    -11.288 &    -12.803   \\
   3.5 &   3.00 &   0.001 &     -6.736 &    -16.498 &     -5.714 &    -10.138 &     -9.551 &    -10.897   \\
\end{tabular}
\end{center}
\end{table}
\begin{table}
\begin{center}
\footnotesize\begin{tabular}{|ccc|cccccc|} \hline\hline
$log\rho Y_{e}$ & $T_{9}$ & $E_{f}$ & \multicolumn{2}{|c|}{$^{54}$Fe}& \multicolumn{2}{|c|}{$^{55}$Fe}
& \multicolumn{2}{|c|}{$^{56}$Fe} \\
\cline{4-9}
& & & \multicolumn{1}{c}{$\lambda_{ec}$} &             \multicolumn{1}{|c|}{$\lambda_{pc}$} &
\multicolumn{1}{c}{$\lambda_{ec}$} &                   \multicolumn{1}{|c|}{$\lambda_{pc}$} &
\multicolumn{1}{c}{$\lambda_{ec}$} &                   \multicolumn{1}{|c|}{$\lambda_{pc}$} \\ \hline
   4.5 &   0.50 &   0.464 &    -20.304 &    -96.203 &     -7.969 &    -48.449 &    -43.561 &    -59.852   \\
   4.5 &   1.00 &   0.309 &    -13.203 &    -48.487 &     -7.867 &    -25.717 &    -24.196 &    -30.588   \\
   4.5 &   1.50 &   0.130 &    -10.690 &    -32.359 &     -7.644 &    -17.732 &    -17.461 &    -20.563   \\
   4.5 &   2.00 &   0.044 &     -9.010 &    -24.429 &     -7.050 &    -13.894 &    -13.710 &    -15.694   \\
   4.5 &   2.50 &   0.020 &     -7.718 &    -19.708 &     -6.342 &    -11.663 &    -11.252 &    -12.839   \\
   4.5 &   3.00 &   0.011 &     -6.719 &    -16.515 &     -5.698 &    -10.154 &     -9.534 &    -10.914   \\
\end{tabular}
\end{center}
\end{table}
\begin{table}
\begin{center}
\footnotesize\begin{tabular}{|ccc|cccccc|} \hline\hline
$log\rho Y_{e}$ & $T_{9}$ & $E_{f}$ & \multicolumn{2}{|c|}{$^{54}$Fe}& \multicolumn{2}{|c|}{$^{55}$Fe}
& \multicolumn{2}{|c|}{$^{56}$Fe} \\
\cline{4-9}
& & & \multicolumn{1}{c}{$\lambda_{ec}$} &             \multicolumn{1}{|c|}{$\lambda_{pc}$} &
\multicolumn{1}{c}{$\lambda_{ec}$} &                   \multicolumn{1}{|c|}{$\lambda_{pc}$} &
\multicolumn{1}{c}{$\lambda_{ec}$} &                   \multicolumn{1}{|c|}{$\lambda_{pc}$} \\ \hline
   5.5 &   0.50 &   0.598 &    -19.338 &    -97.550 &     -6.959 &    -49.797 &    -42.214 &    -61.200   \\
   5.5 &   1.00 &   0.528 &    -12.201 &    -49.592 &     -6.844 &    -26.822 &    -23.091 &    -31.694   \\
   5.5 &   1.50 &   0.423 &     -9.742 &    -33.344 &     -6.689 &    -18.718 &    -16.475 &    -21.549   \\
   5.5 &   2.00 &   0.295 &     -8.389 &    -25.063 &     -6.430 &    -14.528 &    -13.076 &    -16.329   \\
   5.5 &   2.50 &   0.180 &     -7.400 &    -20.031 &     -6.026 &    -11.986 &    -10.929 &    -13.162   \\
   5.5 &   3.00 &   0.110 &     -6.556 &    -16.680 &     -5.536 &    -10.320 &     -9.369 &    -11.079   \\
\end{tabular}
\end{center}
\end{table}
\begin{table}
\begin{center}
\footnotesize\begin{tabular}{|ccc|cccccc|} \hline\hline
$log\rho Y_{e}$ & $T_{9}$ & $E_{f}$ & \multicolumn{2}{|c|}{$^{54}$Fe}& \multicolumn{2}{|c|}{$^{55}$Fe}
& \multicolumn{2}{|c|}{$^{56}$Fe} \\
\cline{4-9}
& & & \multicolumn{1}{c}{$\lambda_{ec}$} &             \multicolumn{1}{|c|}{$\lambda_{pc}$} &
\multicolumn{1}{c}{$\lambda_{ec}$} &                   \multicolumn{1}{|c|}{$\lambda_{pc}$} &
\multicolumn{1}{c}{$\lambda_{ec}$} &                   \multicolumn{1}{|c|}{$\lambda_{pc}$} \\ \hline
   6.5 &   0.50 &   0.905 &    -18.247 &       -100 &     -5.715 &    -52.894 &    -39.118 &    -64.298   \\
   6.5 &   1.00 &   0.880 &    -11.098 &    -51.362 &     -5.650 &    -28.592 &    -21.321 &    -33.464   \\
   6.5 &   1.50 &   0.837 &     -8.634 &    -34.735 &     -5.549 &    -20.108 &    -15.085 &    -22.939   \\
   6.5 &   2.00 &   0.777 &     -7.299 &    -26.276 &     -5.354 &    -15.741 &    -11.863 &    -17.541   \\
   6.5 &   2.50 &   0.701 &     -6.406 &    -21.081 &     -5.063 &    -13.035 &     -9.880 &    -14.211   \\
   6.5 &   3.00 &   0.612 &     -5.738 &    -17.524 &     -4.746 &    -11.164 &     -8.525 &    -11.923   \\
\end{tabular}
\end{center}
\end{table}
\begin{table}
\begin{center}
\footnotesize\begin{tabular}{|ccc|cccccc|} \hline\hline
$log\rho Y_{e}$ & $T_{9}$ & $E_{f}$ & \multicolumn{2}{|c|}{$^{54}$Fe}& \multicolumn{2}{|c|}{$^{55}$Fe}
& \multicolumn{2}{|c|}{$^{56}$Fe} \\
\cline{4-9}
& & & \multicolumn{1}{c}{$\lambda_{ec}$} &             \multicolumn{1}{|c|}{$\lambda_{pc}$} &
\multicolumn{1}{c}{$\lambda_{ec}$} &                   \multicolumn{1}{|c|}{$\lambda_{pc}$} &
\multicolumn{1}{c}{$\lambda_{ec}$} &                   \multicolumn{1}{|c|}{$\lambda_{pc}$} \\ \hline
   7.5 &   0.50 &   1.705 &    -14.323 &       -100 &     -4.113 &    -60.957 &    -31.105 &    -72.360   \\
   7.5 &   1.00 &   1.693 &     -8.917 &    -55.464 &     -4.079 &    -32.694 &    -17.239 &    -37.565   \\
   7.5 &   1.50 &   1.675 &     -6.771 &    -37.550 &     -4.020 &    -22.923 &    -12.281 &    -25.754   \\
   7.5 &   2.00 &   1.648 &     -5.596 &    -28.473 &     -3.903 &    -17.937 &     -9.674 &    -19.738   \\
   7.5 &   2.50 &   1.614 &     -4.834 &    -22.923 &     -3.711 &    -14.877 &     -8.044 &    -16.053   \\
   7.5 &   3.00 &   1.573 &     -4.284 &    -19.139 &     -3.479 &    -12.778 &     -6.916 &    -13.537   \\
\end{tabular}
\end{center}
\end{table}
\begin{table}
\begin{center}
\footnotesize\begin{tabular}{|ccc|cccccc|} \hline\hline
$log\rho Y_{e}$ & $T_{9}$ & $E_{f}$ & \multicolumn{2}{|c|}{$^{54}$Fe}& \multicolumn{2}{|c|}{$^{55}$Fe}
& \multicolumn{2}{|c|}{$^{56}$Fe} \\
\cline{4-9}
& & & \multicolumn{1}{c}{$\lambda_{ec}$} &             \multicolumn{1}{|c|}{$\lambda_{pc}$} &
\multicolumn{1}{c}{$\lambda_{ec}$} &                   \multicolumn{1}{|c|}{$\lambda_{pc}$} &
\multicolumn{1}{c}{$\lambda_{ec}$} &                   \multicolumn{1}{|c|}{$\lambda_{pc}$} \\ \hline
   8.5 &   0.50 &   3.547 &     -1.868 &       -100 &     -1.621 &    -79.532 &    -12.981 &    -90.935   \\
   8.5 &   1.00 &   3.542 &     -1.834 &    -64.782 &     -1.610 &    -42.012 &     -8.044 &    -46.883   \\
   8.5 &   1.50 &   3.534 &     -1.781 &    -43.796 &     -1.599 &    -29.169 &     -6.114 &    -32.000   \\
   8.5 &   2.00 &   3.521 &     -1.713 &    -33.193 &     -1.580 &    -22.658 &     -5.020 &    -24.458   \\
   8.5 &   2.50 &   3.506 &     -1.631 &    -26.735 &     -1.545 &    -18.690 &     -4.291 &    -19.866   \\
   8.5 &   3.00 &   3.487 &     -1.540 &    -22.353 &     -1.490 &    -15.993 &     -3.757 &    -16.752   \\
\end{tabular}
\end{center}
\end{table}
\begin{table}
\begin{center}
\footnotesize\begin{tabular}{|ccc|cccccc|} \hline\hline
$log\rho Y_{e}$ & $T_{9}$ & $E_{f}$ & \multicolumn{2}{|c|}{$^{54}$Fe}& \multicolumn{2}{|c|}{$^{55}$Fe}
& \multicolumn{2}{|c|}{$^{56}$Fe} \\
\cline{4-9}
& & & \multicolumn{1}{c}{$\lambda_{ec}$} &             \multicolumn{1}{|c|}{$\lambda_{pc}$} &
\multicolumn{1}{c}{$\lambda_{ec}$} &                   \multicolumn{1}{|c|}{$\lambda_{pc}$} &
\multicolumn{1}{c}{$\lambda_{ec}$} &                   \multicolumn{1}{|c|}{$\lambda_{pc}$} \\ \hline
   9.5 &   0.50 &   7.583 &      1.215 &       -100 &      0.821 &       -100 &      0.425 &       -100   \\
   9.5 &   1.00 &   7.581 &      1.217 &    -85.136 &      0.821 &    -62.366 &      0.429 &    -67.238   \\
   9.5 &   1.50 &   7.577 &      1.221 &    -57.381 &      0.817 &    -42.754 &      0.436 &    -45.585   \\
   9.5 &   2.00 &   7.571 &      1.226 &    -43.398 &      0.811 &    -32.863 &      0.447 &    -34.663   \\
   9.5 &   2.50 &   7.564 &      1.232 &    -34.917 &      0.810 &    -26.871 &      0.461 &    -28.047   \\
   9.5 &   3.00 &   7.555 &      1.240 &    -29.188 &      0.815 &    -22.827 &      0.479 &    -23.587   \\
\end{tabular}
\end{center}
\end{table}
\begin{table}
\begin{center}
\footnotesize\begin{tabular}{|ccc|cccccc|} \hline\hline
$log\rho Y_{e}$ & $T_{9}$ & $E_{f}$ & \multicolumn{2}{|c|}{$^{54}$Fe}& \multicolumn{2}{|c|}{$^{55}$Fe}
& \multicolumn{2}{|c|}{$^{56}$Fe} \\
\cline{4-9}
& & & \multicolumn{1}{c}{$\lambda_{ec}$} &             \multicolumn{1}{|c|}{$\lambda_{pc}$} &
\multicolumn{1}{c}{$\lambda_{ec}$} &                   \multicolumn{1}{|c|}{$\lambda_{pc}$} &
\multicolumn{1}{c}{$\lambda_{ec}$} &                   \multicolumn{1}{|c|}{$\lambda_{pc}$} \\ \hline
  10.5 &   0.50 &  16.310 &      3.458 &       -100 &      3.313 &       -100 &      3.210 &       -100   \\
  10.5 &   1.00 &  16.309 &      3.458 &       -100 &      3.314 &       -100 &      3.210 &       -100   \\
  10.5 &   1.50 &  16.307 &      3.459 &    -86.714 &      3.314 &    -72.087 &      3.211 &    -74.918   \\
  10.5 &   2.00 &  16.304 &      3.459 &    -65.406 &      3.314 &    -54.870 &      3.211 &    -56.671   \\
  10.5 &   2.50 &  16.301 &      3.460 &    -52.530 &      3.315 &    -44.485 &      3.211 &    -45.661   \\
  10.5 &   3.00 &  16.297 &      3.461 &    -43.874 &      3.317 &    -37.514 &      3.211 &    -38.273   \\
\end{tabular}
\end{center}
\end{table}
\begin{table}
\caption{Comparison of electron capture rates on $^{54,56}$Fe using this model (QRPA), TQRPA \cite{Dzh10} and LSSM \cite{Lan00} at selected temperature and density scales.
\label{tab.2}}
\begin{center}
\footnotesize\begin{tabular}{cccccc} & $log\rho Y_{e}$ & $T_{9}$ & $\lambda_{ec}^{QRPA}$ & $\lambda_{ec}^{TQRPA}$ & $\lambda_{ec}^{LSSM}$
\\\hline
$^{54}Fe$ & & & & &\\
& 7  & 1   & 4.07E-11 & 2.11E-13 & 1.33E-11 \\
& 7  & 1.5 & 1.21E-08 & 4.16E-08 & 3.48E-09 \\
& 7  & 2   & 2.54E-07 & 1.04E-06 & 7.78E-08 \\
& 7  & 3   & 7.87E-06 & 2.44E-05 & 3.30E-06 \\
& 7  & 5   & 2.64E-04 & 5.52E-04 & 1.90E-04 \\
& 7  & 10  & 3.08E-02 & 6.27E-02 & 3.10E-02 \\
\end{tabular}
\end{center}
\end{table}

\end{document}